\documentclass[usegraphicx,aps,jmp,reprint]{revtex4-1}

\def\pmc{{c^4\over{16\pi G}}}

\def\a0{$a_0$}
\begin{document}
\title[MOND and LI Violation]{Hiding Lorentz Invariance Violation
with MOND}

\author{R.H. Sanders}
\affiliation{Kapteyn Astronomical Institute,
P.O.~Box 800,  9700 AV Groningen, The Netherlands}
\date{\today}

\begin{abstract}
Ho\v{r}ava-Lifshitz gravity is an attempt to construct
a renormalizable theory of gravity by breaking the
Lorentz Invariance of the gravitational action
at high energies.  The underlying principle
is that Lorentz Invariance is an
approximate symmetry and its violation by gravitational
phenomena is somehow hidden to present limits of observational
precision.  Here I point out that a simple modification of the
low energy limit of Ho\v{r}ava-Lifshitz
gravity in its non-projectable form  
can effectively camouflage the presence of a preferred
frame in regions where the Newtonian gravitational field
gradient is higher than $cH_0$;  this modification 
results in the phenomenology of MOND at lower accelerations.
As a relativistic theory of MOND this modified Ho\v{r}ava-Lifshitz
theory presents several advantages over its predecessors.
\end{abstract}

\maketitle

\section{Introduction}

A recent theoretical motivation for a universal preferred frame 
is provided by a modern attempt to construct a renormalizable
quantum theory of gravity.  It has been known for some time that
the presence of higher derivatives of the metric tensor
in the field equations can make the theory renormalizable
\cite{stelle}.  However, in a covariant theory, higher spatial 
derivatives also mean higher time derivatives and such theories 
tend to be unstable.  To solve this problem,
Ho\v{r}ava \cite{hor1} has proposed that Lorentz Invariance 
(LI) is broken at high energies by an additional geometrical
structure, a preferred space-like foliation that
splits space from time.  Then it is possible to construct
a renormalizable theory with higher spatial 
derivatives while maintaining
only two time derivatives.  Although
the higher derivative terms become negligible at low
energies, the preferred frame is fundamental and present
at all energies;  it is {\it {assumed}} that the LI violating
terms in the action become negligible at low energies in order
to satisfy phenomenological constraints. 
The LI violation leads to an extra scalar degree of freedom
which in the original ``projectable" Ho\v{r}ava-Lifshitz Gravity (HLG)
presents problems in principle.
Blas, Pujolas and Sibiryakov \cite{bps10, bps11}
(BPS) have shown that these problems are eased
by taking a particular form of Ho\v{r}ava-Lifshitz gravity, the so-called
non-projectable form in which an additional term,
second order in the preferred frame, is added to the action to
provide stability.  But the central
point remains: renormalization in the context of HLG
requires that we live in a Universe in which  
LI is only an approximate or apparent symmetry; the 
Universe contains a fundamental preferred frame that is
somehow hidden to present experimental accuracy.  My 
purpose here is to consider how this camouflage may
be accomplished in terms of a simple modification
of low energy HLG -- a modification leading to MOND phenomenology.

I outline the structure of an acceleration-based 
modification of Ho\v{r}ava-Lifshitz gravity which may be viewed as
a special case of generalized modified
vector-tensor or Einstein-Aether (EA) theories 
\cite{jacmad,jac10,zfs}.  I describe its advantages as a relativistic
theory of MOND and
stress that such a theory can be consistent
with local and cosmological constraints on GR including preferred
frame effects, the absence of Cerenkov losses on high-energy
cosmic rays, and nucleosynthetic limits on the cosmological
value of $G$.  The theory provides enhanced gravitational lensing
about distant astronomical objects (as though by dark matter)
without the construction of a second disformally related
metric.  Therefore, independent of the connection with
HLG, this modified EA theory, as a relativistic theory of MOND,
has relatively few additional ad hoc elements.

\section{Background}

In Ho\v{r}ava-Lifshitz gravity LI is broken at high energies
by the presence of a preferred foliation
of 3-D surfaces in space-time.  In the BPS version, the 
splitting is dynamical; i.e., level
surfaces of a dynamical scalar function $C$ define the foliation
(BPS call this the ``khronon" field). The foliation persists
to low energies but presumably with LI breaking terms strongly
suppressed.

It is useful to define a unit vector which is the normalized
khronon field gradient,
$$A_\mu = {C_{,\mu}\over \sqrt{-g^{\alpha\beta} C_{{,\alpha}} 
{C_{,\beta}}}} \eqno(1)$$  
Then at low energies the modified BPS non-projectable form of HLG may be
written in covariant form as  
$$S_{HLG} = \pmc\int{[R - L_C] \sqrt{-g}d^4 x}\eqno(2)$$
with 
$$L_{C} = (\lambda-1) (\nabla_\mu A^\mu)^2 + \alpha a_\mu a^\mu 
\eqno(3)$$
where $a^\mu =  A^\rho{\nabla_\rho} A^\mu$ is
the acceleration of curves normal to the foliation surface.
The invariant $a_\mu a^\mu$ is the
term added by BPS for stability of the scalar degree of freedom
with $\alpha$ as the dimensionless constant characterizing its 
contribution to the energy-momentum.  

This is recognized as a subclass of EA theories \cite{jac10}
where 
$$L_{EA} = M^{abmn} \nabla_a A_m \nabla_b A_n \eqno(4)$$
with $$M^{abmn} = c_1 g^{ab}g^{mn} +c_2 g^{am}g^{bn} + c_3 g^{an}g^{bm}
  +c_4 A^a A^b g^{mn} \eqno(5)$$
As written here, the only non-zero couplings are $c_2=\lambda -1$
and $c_4=\alpha$.  Given the definition of $A_\mu$ as a 
scalar field gradient, the theory
would appear to contain dangerous higher derivatives of the field $C$.
However, this is avoided when one transforms to the special frame
where $C$ becomes the time coordinate \cite{jac10}.  
Then one may easily show that 
$$A_\mu = \delta_{\mu C} (-g^{CC})^{-{1\over 2}} \eqno(6) $$
and $$a_i = [\ln(N)]_{,i} \,\, \eqno(7)$$
with $N=\sqrt{-g_{CC}}$, the lapse function
in the language of ADM 3+1 formalism.  Here ``$,i$" refers
to spatial derivatives in the preferred frame.  

The first term
in eq.\ 3 involves only time derivatives of the metric, so considering 
the static case, we may take 
$\lambda \approx 1$ (reconsidered below). Then in the preferred frame
the action of the BPS extended HLG may be written simply as
$$S_{HLG} = \pmc\int{\sqrt{-g}[R - \alpha{{N_{,i}N^{,i}}\over {N^2}}]d^4x} 
\eqno(8)$$

In this frame the Einstein equations become
$$R_{ij}-{1\over 2}g_{ij}R = {{8\pi G T_{ij}}\over{c^4}} 
+ \alpha\Bigl[{{N_{,i}N_{,j}\over {N^2}}
     -{1\over 2} g_{ij} {{N_{,k}N^{,k}}\over{N^2}}}\Bigr]\eqno(9)$$
Keeping in mind that in this frame the square of the lapse
is identified with $g_{CC}$ (L.\ Blanchet, private communication), 
the equation for the lapse is given by
$$g^{CC} R_{CC} - {1\over 2} R = {{8\pi G}\over {c^4}} g^{CC} T_{CC}
 +\alpha{{[g^{ij}N_{,i}]_{;j}}\over{N}} - {\alpha\over 2} {{N_{,i}N^{,i}}\over{N^2}}  \eqno(10)$$
One should note that this is the field equation written in the
preferred frame and not in covariant form. Covariant equations
may be found
in references \cite{jac10,lb}.  It is also noteworthy that in this frame the new degree of
freedom, the scalar Khronon field, does not appear explicitly
(it is the time coordinate); 
the LI breaking due to HLG is apparent as revised Einstein equations.

I take small perturbations about Minkowski space of the form
$$g_{ij} = \delta_{ij}(1-2\psi) \,\,\,\,\,\,\,\,\,\, g_{CC} = -(1+2\phi) 
\eqno(11)$$
where the scalars $\phi$ and $\psi$ are the
usual Newtonian potentials. This implies that $N=1+\phi$. 
Then in the static case, 
$$\psi=\phi   \eqno(12)$$
due to the absence of first order contributions to the source
of $G_{ij}$, and
$$2\nabla^2\psi - \nabla\cdot[\alpha\nabla\phi] = 8\pi G \rho/c^2 \eqno(13) $$
where $\alpha$ is included inside the gradient operator because in
the modified version, considered below, it is dependent upon spatial
position in the preferred frame.  Note that this is equivalent
to the two-field non-relativistic theory described by the 
Lagrangian
$$L_{NR} = 4\nabla\phi\cdot\nabla\psi - 2\nabla\psi\cdot\nabla\psi
    - \alpha \nabla\phi\cdot\nabla\phi + {{16\pi G}\over{c^2}} \phi\rho\, \eqno(14)$$

As stressed by several authors \cite{sot,hor11,vis11} the essential difficulty 
concerns the LI violating parameters, $\alpha$ and $\lambda-1$.
There is nothing in the theory that requires these to become
small enough to be consistent with the very tight phenomenological
constraints on deviations from GR (absence of gravitational preferred
frame effects in the Solar System, extragalactic propagation of 
high energy cosmic rays, variations between local and cosmological
values of $G$, binary pulsar constraints).  Below I describe
a modification of HLG which can fit this bill;  a modification
attached to gravitational acceleration rather than energy.

\section{Modified Ho\v{r}ava-Lifshitz gravity}

Modifications of EA theories as possible relativistic 
extensions of MOND have been considered by Zlosnik, Ferreira
and Starkman \cite{zfs} (ZFS) who substitute for this general
vector field Lagrangian (eq.\ 5) a specific function of that
Lagrangian,  $F(L_{EA})$.  It would seem a trivial step to modify HLG
in the same way by taking an appropriate function of
$L_C$ in eq.\ 3. But this would be equivalent to setting
$\lambda-1 \propto \alpha$; here I write a more general
modification by adding two potentials to the HLG Lagrangian, i.e.,
$${L_C}' = L_C + V(\alpha) + U(\lambda') \eqno(15)$$
where $\lambda' = \lambda-1$.  This is equivalent to replacing the
two vector field invariants in eq.\ 3 by separate functions
of these invariants. The parameter 
$\alpha$ is then given by the solution to
$${{dV}\over{d\alpha}} = -\phi_{,i}\phi^{,i}\eqno(16)$$
and $\lambda'$ by $${{dU}\over{d\lambda'}} = -{\sqrt{{g^{CC}}\over {g}}}
{{dg}\over{dC}}\eqno(17)$$ where $g$ is the determinant of the
metric tensor (recall that $C$ is the time coordinate).  

I introduce the modification as added potentials (as in TeVeS)
rather than taking non standard kinetic terms because this leaves
unchanged the form of the LI breaking invariants in the HLG
Lagrangian (eq.\ 3); the kinetic terms are standard.
In any case, there are no underlying principles here;
such ad hoc modifications of scalar or vector field invariants
are not new and have generally been a basis for relativistic
MOND (TeVeS, for example) or dynamical dark energy
(k-essence, for example).

Given the equality of the Newtonian potentials (eq.\ 12), in the weak 
field static case the equation for $\phi$ (eq. 14) becomes
$$\nabla\cdot[\mu({|\nabla\phi/a_0|})\nabla\phi] = 4\pi G\rho\eqno(18)$$
with $\mu = 1-\alpha/2$.  This is recognized as the 
Bekenstein-Milgrom \cite{bekmil} non-relativistic MOND field equation.

For the desired phenomenology the potential must be normalized
by the single new physical constant -- the MOND acceleration parameter
${a_0}^2$.
In the Newtonian limit, where $\alpha \rightarrow 0$,
an appropriate form for the potential would be
$$V(\alpha) = {{{(2^p})}\over {p-1}} \alpha^{(1-p)}[{a_0}^2] \eqno(19) $$ 
where $p>0$ (the special case of $p=1$ corresponds to
a logarithmic potential).  
In the MOND limit where $\alpha\rightarrow 2$
$$V(\alpha) = {2\over 3} (1-\alpha/2)^3 [{a_0}^2]\eqno(20)$$
Note that the natural scale of the potential
is ${a_0}^2$ which would be the approximate value of
any cosmological term (${H_0}^2)$.

The remarkable aspect of Ho\v{r}ava-Lifshitz Gravity is the equality
of the two Newtonian potentials (eq.\ 12).   This implies that the 
relation between
the weak-field force and the deflection of photons is 
identical to that of GR without the construction of two
disformally related metrics, gravitational and physical, 
as in TeVeS \cite{bek04}.  
This fact was first appreciated by ZSF in connection the modified EA theories.
 
Because the potential $V(\alpha)$ may be chosen
such that the theory is arbitrarily close to GR 
in the Solar System, one might expect that the
various post-Newtonian parameters (PPN) may be pushed to within current
experimental accuracy of their GR values.  
But the advantage of considering modified
HLG as a generalized EA theory is that constraints on
on the coefficients, $c_i$, have been worked out for
EA theory \cite{fosjac}.  In particular, the inevitable preferred frame
effects are parameterized by
$${\alpha_1}^{PPN}= -4\alpha\eqno(21)$$
and, in the limit where $\alpha$ is small,
$${\alpha_2}^{PPN} = -5\alpha/2 \eqno(22)$$
For example, taking $V(\alpha)=-{a_0}^2\ln(\alpha)$ 
where $\alpha<<1$ then, at the position of the 
earth in the Solar System,
this gives $\alpha = 2.8\times 10^{-16}$ or 
${\alpha_2}^{PPN}= -7\times 10^{-16}$ in the
neighborhood of the earth, well below the observed 
constraint of $\alpha_2<10^{-7}$.

For time dependent problems, cosmology or wave propagation, 
it is necessary to consider the second parameter
$\lambda' \neq 0$.  A number of the results of
modified EA theory \cite{zuntzea} are directly 
applicable to the model suggested
here but with the restrictions relevant to non-projectable HLG.
For an isotropic homogenous Universe (FRW), $\lambda'$ is
time-dependent and given by the solution to
$$9H^2 = -dU/d\lambda'\eqno(23)$$  That is to say, $\lambda'$
is a function of cosmic epoch (taking $\lambda' \propto \alpha$
in the spirit of the ZFS modification would imply
that the cosmological value of $\alpha$ would be much smaller
in the early universe -- no MOND at earlier epochs).  
Normalizing $U(\lambda')$ by ${a_0}^2$ 
and defining $$K=9{{H^2}\over{c^2 a_0}^2}\eqno(24)$$ one may write
the LI violating term, in the language of modified EA theories,
as $F(K) = \lambda'K$.
Then the modified Friedmann equation given in [14]
becomes 
$$\Bigl[1+3\lambda'\bigl({{d\ln(\lambda'})\over{d\ln(K)}}
+{1\over 2}\bigr)\Bigr]H^2 = {{8\pi G}\over 3} +{U(\lambda')\over 6}
+{V(\alpha)\over 6}\eqno(25)$$ where the cosmological
potentials $U$ and $V$ are special to the present model.

If $\lambda'$ is constant one recovers
the well-known result \cite{cali} 
$$G_c = G_N/(1+3\lambda'/2)\eqno(26)$$
That is to say, the cosmological value
of $G$ is generally less than the local Newtonian 
value.  Observations of the abundances
of light isotopes in the context of 
primordial nucleosynthesis constrain $|G_c/G_N -1|<0.13$.
This places restrictions on $U(\lambda')$.
Taking a generic form $dU/d\lambda' = -\lambda'^{-p}$ we find 
that if $p=2$ then $G_c=G_N$ with no modification to
standard Friedmann expansion (apart from the cosmological
potentials).  If $p=1$, corresponding to a logarithmic
potential ($U(\lambda')\propto -\ln(\lambda')$) 
then $G_c=G_N$ but with a cosmological term
on the order of $({a_0/c})^2$.  Other values of $p$ yield
quintessence.  Therefore, the theory possesses sufficient
flexibility to avoid the nucleosynthetic constraint and to
embody dark energy as well as MOND.  

I stress that the dark energy is not a necessary aspect of 
the theory.  For example, one may set $\gamma=0$ and 
let $\alpha$ be a function of the invariant it multiplies
($N_{,i}N^{,i}/N^2$ in the preferred frame) -- a function 
designed to yield MOND in the low acceleration limit 0.
This would, in the absence of fluctuations, yield a
completely standard cosmology with no cosmological constant
or dark energy, only the gravitational attraction would
be modified.  However, while
dark energy can be fine-tuned away, it would seem to
be a positive attribute that dark energy of the right
magnitude (${a_0}^2$) is most naturally included.

Properly speaking, HLG is a restricted case of
EA theory in which
the vector field $A^\mu$ constrained to be
orthogonal to surfaces of constant $C$ (eq.\ 1); it 
is hypersurface orthogonal \cite{jac10}.  For wave propagation this means that, 
in addition to the usual tensor mode of GR, only one 
longitudinal (scalar) mode can propagate, and not the additional
transverse modes (vector) of the general EA theory. 
 For the hypersurface orthogonal theory considered, 
the usual gravitational
radiation, the tensor mode, propagates at the speed of 
light.  The propagation velocity of the 
scalar mode is
$${c_s}^2 = \Bigl({{\lambda'}\over{3\lambda'+2}}\Bigr) 
\Bigl({{2-\alpha}\over\alpha}\Bigr) \eqno(27)$$
Independently of the value of $\lambda'$ the 
scalar waves must propagate at a velocity below
the speed of light in the cosmological limit where $\alpha\rightarrow 2$.  
This would previously have been
considered fatal for the theory because of the perceived
Cerenkov constraints.  But recently Milgrom \cite{moti11} has pointed
out that the near field of a highly relativistic
particle is in the Newtonian regime (acceleration greater than $a_0$)
which implies a stopping
distance due to gravitational Cerenkov comparable
to the Hubble scale;  there is no problem with
Cerenkov losses for theories that approach GR in
the regime of high accelerations.  Therefore, the 
theory remains valid even given subluminal
longitudinal wave propagation.

Note that radiation damping of compact binary pulsar systems
agree with the predictions of GR to within one percent.  This means that
$\alpha,\lambda'<0.01$ \cite{blasan}, a requirement easily met
by the modified HG theory in the high acceleration regime.

\section{Conclusions}

It has been demonstrated \cite{zfs} that modified EA theories
can form a relativistic basis for MOND.  I emphasize
here that this includes non-projectable HLG as a restricted
EA theory.  HLG in no sense implies 
MOND;  the theory must be modified by making the 
the coupling of the additional
BPS invariant ($\alpha$) dependent upon the invariant
(either by adding a potential $V(\alpha)$ 
or taking a specific function of the invariant as in
modified EA theories).

As relativistic generalizations of MOND, these modified
EA or HLG theories are appealing.  There are relatively few
additional parameters apart from $a_0$; in fact the
parameter $\alpha$ defines the
MOND interpolating function and is required to have
definite limits.  Moreover, unlike TeVeS, such theories
inevitably approach GR in the high acceleration limit.
This means that the classical tests of GR can be readily 
satisfied at current levels of accuracy and, most importantly
(following Milgrom),
there is no Cerenkov constraint on the propagation of
energetic cosmic rays.  The theory is consistent with
observations of cosmic gravitational lenses (i.e., additional
deflection of photons by ``phantom" dark matter) without the
ad hoc construction of a disformally related physical metric.

Solar system and galaxy scale phenomenology can be consistent
with this theory but constrain
the form of the potential $V(\alpha)$.  The cosmology is 
standard but with cosmological terms (dark energy), constant
or evolving, of approximately the correct magnitude (${a_o}^2$).
As in all EA theories, the cosmological $G$ may differ
from its local value, and the limits imposed by primordial
nucleosynthesis constrain the form of the potential
$U(\gamma')$.  At present, the potentials are ad hoc;  
there are no a priori considerations
which tell us what these potentials should be.
Issues such as the viable cosmologies and the growth
of fluctuations will be discussed in a later paper.

In a general sense, HLG requires that 
Lorentz invariance is not a fundamental symmetry but
is violated by gravitational 
phenomenology.  But why then is this LI violation 
not evident in the world around us as, for example, observable
gravitational preferred frame effects?  
The MONDian modification provides one possible solution to this
problem: the theory becomes GR to high precision in the
high acceleration environment of the Solar System.  The dynamical
effects of the preferred frame are hidden by the modified
HLG Lagrangian;
MOND phenomenology occurs in the transition in the outer 
regions of galaxies between local
dynamics described by GR and a preferred frame LI violating cosmology. 
This of course is speculative, and there are other possible
theoretical bases for MOND such as Milgrom's BiMOND \cite{moti09}. 
However, the modified HLG is simple and efficient;  it does not
add many new elements, and it is connected to a well-motivated
approach to quantum gravity. 

The proposed theory is a viable relativistic extension
of MOND, but there remain a number of issues to consider before this proposal could be taken as
a viable extension of HLG, primarily connected with the MOND
{\it Ansatz}:  the variability of $\alpha$ and $\lambda'$
through the 
addition of potentials $V(\alpha)$ and $U(\lambda')$.  
Since the possibility of power-counting renormalization is
the motivation for breaking LI in Ho\v{r}ava-Lifshitz gravity,
then one must ask if the assumption of acceleration-dependent 
couplings vitiates this attribute.  My intuition is that it does not.
Power-counting renormalization in this context requires that
the operators in the action have dimension (number of
spatial derivatives) of no more than six.  This is certainly
true of the $\alpha$ term in the MOND limit.  This
modification would seem only to affect the low energy properties
of the theory.  Questions of
stability and/or ghosts should also be reconsidered, as well
as the strong coupling limit, the energy above which
the higher order terms in the full HLG action,
become dominant.  The possibility that 
$\alpha$ and/or $\lambda'$ can be a dynamical fields might
be considered;  perhaps long wavelength 
oscillations in such fields could play a role
as cosmic dark matter.

\begin{acknowledgments}
I am very grateful to Moti Milgrom for many insightful
comments throughout this work and Luc Blanchet for a helpful
comment on the the derivation of the field equations. 
\end{acknowledgments}

\end{document}